\documentclass[12pt]{article}
\usepackage{lineno}

\usepackage[labelformat=empty, labelsep=space, singlelinecheck=false, skip=3pt]{caption}
\usepackage{afterpage}
\usepackage{rotating}
\usepackage{comment}

\makeatletter
\def\hlinewd#1{%
\noalign{\ifnum0=`}\fi\hrule \@height #1 %
\futurelet\reserved@a\@xhline}
\makeatother

\begin{document}
\pagestyle{plain}

\title{\bf Complementary Theory of Evolutionary Genetics}

\author{ Xiaoqiu Huang \\
Department of Computer Science \\
Iowa State University \\
Ames, Iowa 50011, USA }

\date{}
\maketitle

\noindent
Corresponding Author:

\noindent
Xiaoqiu Huang

\noindent
Email address: xqhuang@iastate.edu


\pagebreak

\begin{abstract}
This theory seeks to define species and to explore evolutionary forces and
genetic elements in speciation and species maintenance.
The theory explains how speciation and species maintenance are caused
by natural selection acting on non-Mendelian and Mendelian variation, respectively.
The emergence and maintenance of species
as groups of populations are balanced by evolutionary forces including
complementary mechanisms of gene flow within and between populations 
at population-specific rates:
sexual and asexual reproduction, recombining and nonrecombining genome regions,
vertical and horizontal DNA transfer, and transposon proliferation and control.
While recombining genome regions carry conserved genes and are subjected to meiotic recombination,
nonrecombining genome regions carry accessory genes and are not subjected to such structural restrain.
Sexual reproduction, vertical DNA transfer, recombining genome regions and transposon control
keep species in existence by maintaining recombining chromosome number and structure,
while asexual reproduction, horizontal DNA transfer,
nonrecombining genome regions and transposon proliferation help species emerge
by promoting reproductive isolation and changes in chromosome number and structure.
The theory is based on the analysis of the genome sequences of isolates
in the {\em Fusarium oxysporum} complex.
The rate of horizontal supernumerary chromosome transfer in this complex
was estimated to be $0.1$ per genome per year.
\end{abstract}


\pagebreak


\section*{Introduction}
Darwin (1859) explained the role of natural selection in the origin of species,
and Mendel (1901) discovered the mechanism of inheritance of traits.
Fisher, Haldane and Wright in the 1920s and 1930s developed mathematical models of evolution
as a change in the frequency of gene variants over time.
Muller (1932) predicted that asexual populations accumulate irreversible deleterious mutations.
Dobzhansky (1970) proposed that chromosome translocation contributes to the birth of new species,
and Franchini et al. (2020) suggested that chromosomal rearrangements
play an important role in speciation.
Coluzzi (1982) proposed a chromosomal speciation model of suppressed recombination.
Growing evidence supports a significant role of selfish genetic elements in eukaryotic speciation
(Werren 2011).
Mayr (1942) emphasized reproduction isolation in the concept of species;
reproductive isolation is the critical factor behind 
the emergence of new eukaryotic populations of organisms that reproduce sexually.
Reproduction isolation was linked with variation in
{\em Drosophila} Y chromosome, a nonrecombining genome region (Hafezi et al. 2020).
Asexual reproduction is associated with polyploidy in both plants and animals
(Herben et al. 2017).
Asexual reproduction can be viewed as a genetic form of reproductive isolation
that is common in some groups of eukaryotes such as fungi.
An intriguing question is whether asexual reproduction is the critical factor behind
the emergence of new eukaryotic populations of organisms that reproduce
both sexually and asexually.
Is there a concise genetic definition of eukaryotic populations of organisms
that reproduce asexually most of the time?
What are the nature and rate of gene flow within such populations?
What kinds of changes in genome structure occur in such populations?
In this study, we attempted to collect evidence for these questions in
a complex group of fungal populations called the {\em Fusarium oxysporum} speices complex,
and propose a theory based on the evidence to improve the current theory of evolutionary genetics.




The availability of genome sequence data presents a historic
opportunity to address classical questions in evolutionary genetics:
the evolutionary consequences of changes in chromosome number and structure (Peichel 2017).
Extensive amounts of genome sequence data are available
for isolates in the {\em F. oxysporum} species complex,
which holds the potential to understand the role of horizontal chromosome
transfer in evolution (Kistler et al. 2013).
The {\em F. oxysporum} species complex contained pathogenic fungal populations
for a large number and a wide range of hosts from plants to animals including humans,
where each population had a narrow host range (van Dam et al. 2016).
This implies that a large number of populations were present in the complex.
Although no sexual cycle was observed in the complex,
mating-type loci cloned from this complex was functional in a close sexual relative
(Arie et al., 2000).

The genome of the individual or isolate in the complex
was composed of the core genome containing single-copy genes
(except ribosomal DNA (rDNA) genes)
that were conserved among the isolates in the complex and
the supernumerary genome containing accessory genes that
were present only in certain isolates and that could have multiple copies in the genome
(Covert 1998; Ma et al. 2010; Rep and Kistler, 2010).
The supernumerary genome constituted nearly 40\% of the genome, based on
the difference in genome size between {\em F. oxysporum} and {\em F. graminearum}.
Rates of single nucleotide polymorphisms (SNPs) in conserved genes between isolates were mostly 0-3\%.
The core genome was distributed among core chromosomes, while
the supernumerary genome was distributed among supernumerary chromosomes
and specific regions (e.g. the ends) of core chromosomes.
The supernumerary genome evolved more rapidly than the core genome 
(Raffaele and Kamoun 2012; Croll and McDonald 2012; Dong et al. 2015; Huang et al. 2016).
Supernumerary chromosomes have been shown in {\em in vitro} experiments to
transfer between vegetatively incompatible isolates or to
transfer from a pathogenic isolate to a non-pathogenic isolate in asexual filamentous fungi
(He et al. 1998; Akagi et al. 2009; Ma et al. 2010; Vlaardingerbroek et al. 2016a; van Dam et al. 2017).
In {\em F. oxysporum}, phylogenetic studies suggest horizontal transfer of supernumerary chromosomes
and supernumerary effector genes (van Dam et al. 2016; Fokkens et al. 2018),
and supernumerary chromosomes are likely acquired
by horizontal transfer through vegetative fusion of hyphae (Eschenbrenner et al. 2020).
Several high-quality genome assemblies contained 11 core chromosomes,
one or more supernumerary chromosomes, and sometimes keeper chromosomes,
which contained large supernumerary regions fused with large core regions
of the two smallest core chromosomes that could split into two segments through fission.
Chromosome rearrangements generated supernumerary chromosomes in the wheat blast
fungus (Langner et al. 2021), and supernumerary regions of core chromosomes 
in the {\em Verticillium} wilt fungus were thought to be acquired horizontally (Huang 2014).
Below we present a general theory, a particular model of evolution
for the {\em F. oxysporum} complex, and evidence for the model
from the analysis of sequencing and genomic data in the complex.

\section*{Results}

\subsection*{Theory}

This theory explains how species maintenance is caused
by natural selection acting on Mendelian variation in gene structure,
and how speciation is caused by natural selection acting on non-Mendelian variation
in chromosome number and structure.
The emergence and maintenance of species
as groups of populations are balanced by evolutionary forces including
complementary mechanisms of gene flow within and between populations 
at population-specific rates:
sexual and asexual reproduction, recombining and nonrecombining genome regions,
vertical and horizontal DNA transfer, and transposon proliferation and control.
Nonrecombining genome regions include B and sex chromosomes in plants and animals, and
supernumerary chromosomes (also called dispensable, lineage-specific, or accessory chromosomes) in fungi.
While core chromosomes carry conserved genes and are subjected to meiotic recombination,
nonrecombining genome regions carry accessory genes and are not subjected to such structural restrain.
Sexual reproduction, vertical DNA transfer, recombining genome regions and transposon control
play major roles in maintaining chromosome number and structure,
while asexual reproduction, horizontal DNA transfer
nonrecombining genome regions and transposon proliferation are main genetic factors behind
reproductive isolation and changes in chromosome number and structure.
Nonrecombining genome regions are enriched in genes involved
in genome dynamics, adaptation to environments and reproductive isolation,
where some of those genes arise by horizontal gene transfer,
which is an ongoing evolutionary force during asexual reproduction.
Thus, eukaryotic populations,
especially populations of organisms that reproduce both sexually and asexually,
emerge and adapt by undergoing changes more frequently in nonrecombining genome
structure than in recombining chromosome structure.
The theory is based on a model of evolution for populations in
the {\em F. oxysporum} complex. The model was formulated by studying
changes in chromosome number and structure within and between populations in the complex.
In this complex, for example,
the generation and transfer of supernumerary chromosome structural variants
alongside the formation of population-specific subtelomeric palindromes at the ends of chromosomes
allow pathogenic fungal populations to emerge and evolve during asexual reproduction.
Some of those populations contained
fusions between core and supernumerary chromosomes,
as well as translocations between core chromosomes, which could be
potential barriers to meiotic recombination.
The rate of horizontal supernumerary chromosome transfer
was $1/y$ per genome per year, where $y$ was
the number of years for the isolate to acquire population-specific subtelomeric palindromes
at the ends of its chromosomes for the first time.
The parameter $y$ was estimated to be less than 10 years.

\subsection*{Model of evolution for populations in the {\em F. oxysporum} species complex}

Variation in chromosome structure is a major driver of divergence and speciation, and
stability in chromosome structure is a major keeper of species in existence.
In this sense, speciation and species maintenance are in conflict,
and the evolutionary and molecular mechanisms of speciation and species maintenance
are also in conflict or complementary.
For example, selfish genetic elements are in conflict with other genes
in the eukaryotic genome (Werren 2011).
Possible complementary modes and mechanisms include
sexual and asexual reproduction, recombination and nonrecombination,
vertical and horizontal DNA transfer, core and supernumerary chromosome, 
strong and weak selection, rapid and slow gene flow,
rapid and slow mutation, transposon proliferation and control,
change in the frequency of gene variants and chromosome structural variants,
variation and stability in chromosome number,
genes in single and multiple copies,
chromosomes in single and multiple copies,
genes with benefits in general and particular environments,
large and small genomes,
large and small numbers of individuals,
intact and mutated subtelomeres, and
and haploidy, diploidy and polyploidy.
Most importantly, speciation and species maintenance are strongly affected by the mode of reproduction.
Sexual reproduction is effective at maintaining species by preserving
their chromosome structure through recombination, but less efficient
for variation in chromosome structure to occur.
Because of lack of recombination, asexual reproduction is efficient
for variation in chromosome structure to occur,
but less effective at maintaining species by preserving their chromosome structure.
The weaknesses in sexual and asexual reproduction are
compensated by using mixed modes of reproduction and/or other complementary mechanisms.
In general, speciation and species maintenance are balanced
by those complementary modes and mechanisms at appropriate frequencies and rates.


We explain how speciation and species maintenance
in the {\em F. oxysporum} species complex
were balanced by complementary modes and mechanisms at appropriate frequencies and rates.
The complex was estimated to have existed for at least 10 thousand years.
In this complex, two types of genome were formed to maintain
the complex and to allow different populations to adapt to diverse environments
(Croll and McDonald, 2012; Raffaele and Kamoun, 2012).
The core genome was intended to carry genes with benefits
in all environments, and the supernumerary genome to carry genes with benefits
in particular environments.
The core genome was conserved among isolates,
while the supernumerary genome was highly variable among isolates.
The stability in the core genome was maintained by strong purifying selection
as well as infrequent sexual reproduction, during which
the proliferation of transposable and repetitive elements
was controlled by Repeat Induced Point Mutation (RIP)
(Cambareri et al. 1989; Gladyshev 2017).
The infrequent sexual reproduction offered little time
for a change in the frequency of conserved gene variants to occur in the complex.
The variation in the supernumerary genome was promoted
by the frequent horizontal transfer of supernumerary chromosomes
carrying transposons within and between populations,
where the rate of horizontal transfer was more frequent than sexual reproduction.
Haploidy was the predominant state in this complex, with
single-copy conserved genes (except for rDNA genes) in core chromosomes,
where deleterious mutations in conserved genes were removed more effectively than
in two-copy genes, since their effects were not shielded (Orr and Otto 1994).
On the other hand, genes in supernumerary chromosomes were under weaker purifying
selection and could be present in multiple copies,
and differences in supernumerary chromosome number
and structure were prevalent among isolates.
Selection plays a major role in removing supernumerary chromosomes
with deleterious mutations from the complex
and in increasing the frequency of beneficial supernumerary chromosomes in the complex.
The construction of accurate phylogenetic trees from core chromosomes of
isolates across the complex (Achari et al. 2020; Fokkens et al. 2020)
revealed insignificant flow of core chromosomes in the complex either vertically or horizontally.



The genomes of isolates in the complex contained
the hallmarks of RIP in the form of C-T and G-A mutations in and around
repetitive elements (except rDNA repeats),
which is regarded as the signature of past sexual reproduction,
since RIP occurs during sexual cycles.
Because sexual reproduction in the complex was infrequent and short,
it may be effective in maintaining chromosome number and structure
through recombination and in controlling transposon proliferation through RIP,
but it may not be able to contribute significantly
to generating recombinations of gene variants.
Therefore, most of the population adaptation and divergence in the complex
occurred during long periods of asexual reproduction.

All core and supernumerary and keeper chromosomes were flanked on both sides
by inverted copies of a population-specific subtelomeric element, called a subtelomeric palindrome.
All isolates with subtelomeric palindromes of the same kind
belonged to the same population, which persisted on
specific hosts or in specific environments.
One reason why all periods of sexual reproduction were short
is that isolates undergoing sexual reproduction could not
persist on any hosts or in any environments for a long time
because their host- or environment-specific supernumerary genomes
were all inactivated by RIP during sexual cycles.
The complex contained a diverse set of supernumerary chromosomes
distributed in a large number of populations.
The supernumerary genome was enriched in transposable and repetitive elements,
pathogenicity genes and HET (heterokaryon incompatibility) domain genes
(Paoletti and Clav\'{e} 2007; Vlaardingerbroek et al. 2016b).



The lineage for each isolate was comprised of long periods of asexual reproduction
and short periods of sexual reproduction. 
During a period of sexual reproduction,
the duplicated regions of the supernumerary genome were inactivated
by RIP into AT-rich regions that contained no functional genes;
such duplicated regions included transposons and subtelomeric palindromes.
At the beginning of a period of asexual reproduction,
in order to survive on a host or in an environment,
the lineage had to reconstruct its supernumerary genome quickly by horizontally acquiring
an intact supernumerary chromosome with a population-specific subtelomeric palindrome
and duplicating the palindrome at the end of each core chromosome.
During the period of asexual reproduction, within the nucleus,
gene exchanges between supernumerary chromosomes and core chromosomes occurred
through the homology of the subtelomeric palindromes at the ends of
these chromosomes, and transposons along with pathogenicity genes moved from supernumerary chromosomes
to AT-rich regions of core chromosomes.

Gene flow manifested in the form of the horizontal transfer of supernumerary chromosomes
within and between populations during asexual reproduction.
The horizontal transfer of a supernumerary chromosome
within a population led to isolates with two or more structurally different
copies of the supernumerary chromosome. These copies underwent
chromosome rearrangements so that deleterious variants were lost and beneficial variants
became more prevalent within the population.
When an isolate from one population came into contact with an isolate
from another population, only supernumerary chromosomes might move from
one isolate to the other; the core chromosomes in
one isolate could be separated from those in the other isolate
based on the differences in the sequences of their subtelomeric palindromes,
which was one of the reasons why all core chromosomes in an isolate were flanked on
both sides by subtelomeric palindromes of its population-specific type.
Note that phylogenetic trees of isolates in the complex were accurately constructed
on sequences from core chromosomes, indicating that the extent to which 
core chromosomes from different populations were mixed through horizontal transfer
during a long period of asexual reproduction was minimal,
and also indicating the absence of extensive meiotic recombination in this complex.
This absence of extensive meiotic recombination menas that
the evolution of this species complex was reflected to a lesser extent
through change in conserved gene frequency.
Instead, the horizontal transfer of supernumerary chromosomes
is proposed as a major driver of the evolution of this species complex
through change in genome structure.
Gene flow within and between populations during asexual reproduction
was controlled by over 100 HET domain genes.

A new population of pathogenic isolates could emerge for a host through
the horizontal transfer of a supernumerary chromosome as follows.
First, the host (called $a$) developed resistance to a population $A$ of pathogenic isolates.
Then a new supernumerary chromosome emerged from another population $B$ (for a different host $b$)
that contained multiple supernumerary and keeper chromosomes.
Next the new supernumerary chromosome with the $B$-specific subtelomeric palindrome
arrived in isolates of population $A$ through horizontal transfer, and 
subsequently underwent subtelomeric palindrome changes from type $B$ to type $A$.
After that, the old and new supernumerary chromosomes in population $A$
underwent gene exchanges so that a resulting supernumerary chromosome had a new subtelomeric palindrome
different from that of population $A$ and could cause disease to host $a$.
Finally, the new population was founded when the resulting supernumerary chromosome
arrived in isolates of a population
and caused the core chromosomes of those isolates to obtain
its subtelomeric palindrome. With asexual reproduction, the new population underwent growth
and expanded by spreading its supernumerary chromosome
to isolates of other populations through horizontal transfer.
The above description was based on the similarity between supernumerary chromosomes
from several real populations, and the observation that the subtelomeric palindrome of
one population were found in a supernumerary chromosome in another population.

During a period of asexual reproduction,
certain deleterious mutations such as nucleotide substitutions
in core chromosomes from some nuclei could be removed
through mitotic recombination with matching core chromosomes from other nuclei
within the individual (Nieuwenhuis and James 2016).
However, when the AT-rich regions of core chromosomes were all taken up
by active transposons, deleterious mutations
to core chromosomes caused by the proliferation of transposons 
in all nuclei within the individual could no longer be fixed through mitotic recombination.
This caused the current period of asexual reproduction to be terminated.
A new short period of sexual cycles was needed
to control the proliferation of transposons by RIP,
which is known to induce massive point mutations in duplicated regions rapidly.
These sexual cycles also maintained the number and structure of core chromosomes
through meiotic recombination.
Because transposons arrived on supernumerary chromosomes during
a period of asexual reproduction,
the composition and transfer rate of supernumerary chromosomes 
affected the rate at which structural variation was generated
and the length of this asexual period.
Note that the length of existence of a pathogenic population
depended more on the availability and susceptibility of its host
than on the inability to purge deleterious mutations during asexual reproduction.
An explanation for this is that the horizontal transfer of supernumerary chromosomes
would allow the population to expand in young asexual populations
if the current asexual populations carrying the population had lasted
for a long period of time and accumulated deleterious mutations.

The rate of horizontal transfer was estimated based on the above model.
At least one horizontal transfer was needed for any asexual lineage
to acquire a population-specific supernumerary chromosome
in order to persist in the environment for the population.
Let $y$ be the longest length in years of asexual lineages in the  
{\em F. oxysporum} complex to acquire their first population-specific supernumerary chromosomes
after sexual reproduction. Then the rate of horizontal
transfer for supernumerary chromosomes between populations
was bounded from below by $1 / y$ per genome per year.
If $y$ was less than 10 years, then the rate of transfer
was greater than $0.1$ per genome per year.
An argument in favor of a small value for the parameter $y$
could be supported by the inability to find natural isolates without intact subtelomeric palindromes.
Let $\lambda$ be the longest length in years of asexual lineages in the  
{\em F. oxysporum} complex.
The rate of transposition for transposons was bounded from below by $\tau /  \lambda$,
where $\tau$ is the minimum number of intact transposons in core chromosomes in any asexual lineage.
If ${\lambda}_i$ was the length in years of the asexual lineage for isolate $i$ with
a population-specific subtelomeric palindrome, then the rate of
horizontal transfer within the population for this lineage was $c_i / {\lambda}_i$ per genome per year,
where $c_i$ was the number copies of a supernumerary chromosome in isolate $i$.

In the complex, dynamic genetic populations were carried by static asexual populations.
As some asexual populations became old, they
passed their genetic populations to young asexual populations through
the horizontal transfer of the supernumerary chromosomes 
in the genetic populations.
Then the old asexual populations underwent sexual cycles
to become young asexual populations. 
A major role of sexual reproduction was to produce young asexual populations
with the correct number and structure of core chromosomes
and with fewest functional transposons so that
those young asexual populations would be able to carry
dynamic genetic populations.
Note that gene duplication, which is an
important mechanism by which evolution occurs, was restricted
to the supernumerary genome.
An explanation for this restriction is that the core genome
went through sexual cycles, which would inactivate all duplicated genes,
so that only the unique core genome could stay intact during sexual cycles.
Supernumerary chromosomes with duplicated genes were transferred from old asexual populations
to young ones, bypassing sexual cycles so that they were not subjected to RIP.
Note that supernumerary chromosomes that remained in isolates undergoing sexual cycles
would be inactivated by RIP and then were lost.
Transposons in supernumerary chromosomes in the wheat blast fungus
lacked signature of RIP (Peng et al. 2019).

\subsection*{Evidence: Population-specific subtelomeres and AT rich regions}

We previously found that in isolates Fol race 3 D11, Forc Forc016, Fom Fom001 and
Focb race 1 160527, core and supernumerary chromosomes were
flanked on both sides by inverted copies of a host- or population-specific subtelomeric element
(Huang 2019). That is, the 5' copy in forward orientation
was highly similar to the 3' copy in reverse orientation, with both
copies located within 10- to 15-kb ends of the chromosome, respectively.
This pair of inverted copies was called a subtelomeric palindrome here.
The finding also held for isolates Fo47, Foc Fo5176 and Focb TR4 UK0001.
The subtelomeric element contained a gene encoding a helicase,
where the reading frame of the gene was free of stop codons (Huang 2019).

We examined AT rich regions in the genomes of the 15 {\em F. oxysporum} isolates
listed in the Materials and Methods.
Of the 15 isolates, 5 isolates
(Fo47, Foc Fo5176, Fol D11, Focb 160527 and Focb TR4 UK0001)
had a high-quality genome assembly; in each of these 5 isolates,
the AT content of each core and supernumerary chromosome was around 52\%.
On the other hand,
the genome assemblies of all the 15 isolates
contained AT rich regions of total sizes ranging from 511.3 to 2,270.1 kb with a mean of 1,028.9 kb,
where an AT rich region was a block of consecutive lines of sequences with
the AT content of each line at 65\% or higher and with the size of the block at 2 kb or larger.


Some AT rich regions were inactivated subtelomeric palindromes
containing helicase genes with many stop codons in their reading frames.
For example, isolate Fo47 chromosome 2 contained a pair of inactivated subtelomeric elements
that were 15 to 16 kb away from the respective chromosome ends, each of which contained
an intact subtelomere with no stop codons in its reading frame.
The 5' subtelomere had an AT content of 73\% over a length of 3.7 kb, and
was 45\% identical over 1313 residues to an ATP-dependent DNA helicase hus2/rqh1 of 1,428 residues
(accession: KAG7001869.1) encoded in isolate Fo5176 supernumerary chromosome 18.
This DNA-protein alignment contained 100 stop codons in the reading frame of the 5' subtelomere.
Similarly, the 3' subtelomere had an AT content of 69\% over a length of 1.4 kb, and
was 37\% identical over 593 residues to the same helicase protein,
with 37 stop codons in its reading frame.
This example suggested that an ancestor of Fo47 chromosome 2 contained inactivated subtelomeres
at both ends and later acquired an intact subtelomere attached to each of its ends,
where the intact subtelomere was located within the new end of length 15 kb.

In another example, isolate Forc Forc016 supernumerary chromosome RC
contained a pair of intact subtelomeric elements, each of whose reading frames
was 90\% identical over 984 residues to the above helicase protein with no stop codons
in the reading frame. This also happened with the intact subtelomeric elements
in the core chromosomes of this isolate. Those similarities could explain the 
origin of the current set of subtelomeric palindromes in this isolate.
Moreover, the isolate contained several inactivated subtelomeric palindromes
that were similar to the helicase protein with many stop codons in their reading frames.
For instance, we found two 3' inactivated subtelomeres that were 91.7\% identical over 4.0 kb,
adjacent to two intact subtelomeres with a nearly perfect match over 13 kb.
A similar case also happened to another two 3' inactivated subtelomeres next
to two intact ones.
These examples suggested that inact subtelomeres were attached to
ancetor chromosome ends containing inactivated subtelomeres.
Lastly, isolate Focb race 1 160527 contig 7 of 4.2 Mb contained a 3' inactivated partial 4.0-kb
subtelomere with an AT content of 68\% that was 54\% identical over 1,276 residues
to the helicase protein with 78 stop codons in the reading frame,
and isolate TR4 UK0001 contig 4 of 5.2 Mb contained an 5' inactivated partial 3.7-kb subtelomere
that was 42\% identical over 1,078 residues
to the helicase protein with 93 stop codons in the reading frame.
Note that these two subtelomeres were located
at 14 kb away from the 3' end and at 21 kb away from the 5' end, respectively.
Those AT regions were evidence for past RIP activities during sexual cycles,
which mutated repetitive elements including subtelomeric palindromes.

Extremely high and variable rates of accessory chromosome loss
were reported in the plant pathogenic fungus {\em Zymoseptoria tritici},
which reproduces both sexually and asexually, with sexual cycles observed.
In {\em Z. tritici} isolate ST00Arg\_1D1a1,
a 3' subtelomere of 11.4 kb of chromosome 12 was 94-95\% homologous
to the subtelomeres of 14 other chromosomes with 99.4\% of 22,679 substitutions
being G-A and C-T mutations, the signature of the RIP process during a sexual cycle.
A longest region with an AT content of $\geq 65\%$ in the {\em Z. tritici} isolate ST00Arg\_1D1a1 was 11 kb,
and that in the {\em F. oxysporum} isolate UK0001 was 149 kb.

\subsection*{Evidence: Gene duplication and structural variation}


We considered gene duplication and structural variation in isolate Fo47,
Foc race 1 isolates Fo5176 and IVC-1, each of which had a high-quality genome assembly.
A hallmark of asexual reproduction in {\em F. oxysporum} pathogenic isolates
was extensive duplication of accessory genes within the genome
(within and between supernumerary chromosomes and the ends of core chromosomes). 
For example, we found more than 46,000 gap-free matches of at least 5 kb in length
and at least 99\% identity in the genome assembly of isolate Fo5176,
and as a comparison, 18 such matches in the non-pathogenic isolate Fo47.
All of these 18 matches in isolate Fo47 were between subtelomeres
and other supernumerary regions. This showed that gene duplication
was limited to accessory genes in isolate Fo47, with all conserved genes
(except for rDNA genes) as single copy genes.
We also checked on matches between the genome assemblies
of isolates Fo47 and Fo5176, and found that all non-unique
matches were between supernumerary regions of the two isolates.
Those comparisons confirmed that all the gene duplications
in isolate Fo5176 were within its supernumerary genome region.

Next we quantified the amount of large-scale structural variations
between Foc race 1 isolates Fo5176 and IVC-1 with respect to the number of SNPs between them.
We used minimap2 (Li 2018) with a stringent assembly-to-reference mapping option (the preset -x asm5 option)
to compute unique alignments of lengths at least 200 kb between the Fo5176 and IVC-1 genome assemblies,
where the minimum alignment length of 200 kb was selected because all duplications
within the Fo5176 assembly were of lengths less than 200 kb.
These unique alignments contained 88 deletion gaps of lengths from 1,146 to 8,662 bp
with a total deletion gap length of 303,479 bp, and 85 insertion gaps of lengths from 1,323 to 12,268 bp
with a total insertion gap length of 335,954 bp.
Here, a deletion (an insertion) gap was composed only of a sequence region from the Fo5176 (IVC-1) assembly,
and the length of the gap was the length of the sequence region.
These large deletion and insertion gaps were likely to be some of the structural variations (SVs)
between the two race 1 isolates with a total of 1,072 SNPs, resulting
in a ratio of the SV total length to the SNP number at the level of 639,433/1,072 = 596.5.
Many of those sequence regions in the gaps were comprised of
transposons present in multiple copies in the genome.
So this approach allowed us to quantify SVs in repetitive regions of
the genome.

To quantify SVs between races 1 and 2 in unique regions of the genome,
we mapped short reads from race 1 isolate IVC-1 and race 2 isolate 58385
onto the genome assembly of race 1 isolate Fo5176 as a reference.
An SV between race 1 isolate IVC-1 and race 2 isolate 58385
was defined as a reference region of length at least 1 kb such that
the read depths for the two isolates at every position of the reference region
consistently indicated the presence of coverage by one isolate (called P)
and the absence of coverage by the other isolate (called A).
Precisely, the following conditions hold at every position x of the region:
$dep(P, x) \geq 0.7 * average\_dep(P)$ and $dep(A, x) \leq 0.05 * average\_dep(A)$,
where $dep(I, x)$ is the read depth of isolate $I$ at position $x$,
and $average\_dep(I)$ is the genome-wide average read depth of isolate $I$.
We found 154 SVs of a total length of 315,107 bp with the presence of coverage only by race 1 isolate IVC-1 
and 108 SVs of a total length of 318,650 bp with the presence of coverage only by race 2 isolate 58385.
The number of SNPs between race 1 isolate IVC-1 and race 2 isolate 58385 was 3,225.
Thus, these numbers resulted in a ratio of the SV total length to the SNP number at the level of
633,757/3225 = 196.5.

\subsection*{Evidence: Fusions between large core and supernumerary regions}

We studied chromosome-level variation in several isolates by comparing their genome assemblies.
Isolates in the species complex are known to carry 11 core chromosomes
and one or more supernumerary chromosomes (Ma et al. 2010).
The genome assembly of Fo47 was composed of 12 chromosomes named 1 through 12
in decreasing order of chromosome sizes, with 11 core and 1 supernumerary (chromosome 7).
The two smallest core chromosomes were less conserved than the other core chromosomes.
By comparing the genome assembly of Fo47 to those of the other isolates,
we found in some of the other isolates that one of the two smallest core chromosomes (homologous
to chromosomes 11 and 12 of Fo47) or a large region ($\geq 900$ kb) of it was fused
with a supernumerary chromosome or a large region of it.

In the genome assembly of Fol race 3 isolate D11,
contig 1 was a result of a fusion between a 1,120-kb core region (as a 5' portion) and
a 4,680-kb supernumerary region (as a 3' portion),
where the 1,112-kb core region was homologous to a 3' portion of core chromosome 12 in isolate Fo47.
Similarly, contig 6 of isolate D11 was composed of a 1,206-kb supernumerary region (as a 5' portion)
and a 929-kb core region (as a 3' portion), with the core region being
homologous to a 3' portion of core chromosome 11 in isolate Fo47.
No such fusion events were detected in the genome assembly of Fol race 2 isolate 4287.

The genome assembly of Focb race 1 isolate 160527 was composed of 12 contigs named 1 through 12.
Contig 2 of 5,885.8 kb was a result of a fusion
between a core chromosome (homologous to core chromosome 11 in isolate Fo47) and
a 3,211-kb supernumerary region (as a 3' portion). This fusion was confirmed by long reads.
Contig 12 was a standalone supernumerary chromosome of 1,261 kb.
All other contigs were core chromosomes based on comparison with the Fo47 chromosomes;
their ends may contain short supernumerary regions ($\leq 200$ kb).

The genome assembly of Focb TR4 isolate UK0001 was made up of 15 contigs,
three of which were of lengths less than 120 kb. Among the remaining 12 contigs,
contig 13 of 1.24 Mb was a standalone supernumerary chromosome,
and contig 14 of 3.74 Mb was fused between a core chromosome (homologous
to core chromosome 12 in isolate Fo47) and a supernumerary region of 1.30 Mb (as a 3' portion).
All other contigs were core chromosomes based on comparison with the Fo47 chromosomes.

We examined a genome assembly of Foc race 1 isolate Fo5176, which was composed of 19
chromosomes named 1 through 19. To find chromosomal fusions and translocations in
this genome assembly, we compared it with that of isolate Fo47.
A first chromosomal difference between the two genome assemblies
involved Fo5176 core chromosome 5 of 5.04 Mb and chromosome 6 of 5.01 Mb
as well as Fo47 core chromosome 2 of 5.61 Mb and chromosome 4 of 4.73 Mb.
To describe this difference, a minor portion was used to refer to a smaller 5' or 3' region
of a chromosome and a major portion to the larger remaining part.
Specifically, a 5' minor portion of Fo5176 chromosome 5 (chromosome 6)
was syntenic to a 3' minor portion of Fo47 chromosome 2 (chromosome 4);
the major portion of Fo5176 chromosomes 5 (chromosome 6) was syntenic
to the major portion of Fo47 chromosome 4 (chromosome 2).
Each of Fo47 core chromosomes 2 and 4 was syntenic to a single contig or scaffold
in each genome assembly of isolates 160527 and UK0001.
This observation suggests an event of reciprocal translocation in the Fo5176 lineage
in which a 5' minor portion (totalling 1.79 Mb) of an ancestor core chromosome
was exchanged with a 5' minor portion (totalling 1.45 Mb) of another ancestor core chromosome.
This exchange was not present in isolates Fo47, 160527 or UK0001.


A second chromosomal difference associated Fo5176 chromosome 4 of 5.26 Mb and
Fo5176 chromosome 13 of 2.80 Mb with Fo47 core chromosome 11 of 2.85 Mb.
A core portion (at positions 0.54 to 1.80 Mb) of Fo5176 chromosome 4 was syntenic to
a portion (at positions 0.57 to 1.66 Mb) of Fo47 core chromosome 11, and
a core portion (at positions 1.68 to 2.69 Mb) of Fo5176 chromosome 11 was syntenic to
a portion (at positions 1.66 to 2.50 Mb) of Fo47 core chromosome 11.
The remaining portions of Fo5176 chromosomes 4 and 11 were mostly supernumerary.
This difference was a result of chromosomal fusions between core
and supernumerary chromosomes or chromosomal regions in the Fo5176 lineage
(Fokkens et al. 2020).


A third difference connected Fo5176 chromosome 10 of 3.19 Mb to Fo47 core chromosome 10
of 2.89 Mb. A 5' minor portion (totalling 0.80 Mb) of Fo5176 chromosome 10
was syntenic to a 5' minor portion of Fo47 core chromosome 10,
and the major portion of Fo5176 chromosome 10 was supernumerary;
the major portion of Fo47 core chromosome 10 was syntenic to the majority of
Fo5176 chromosome 15 of 2.40 Mb.
This finding indicated that Fo5176 chromosome 10 resulted from
a fusion between a core chromosomal region and a supernumerary chromosome
(Fokkens et al. 2020).


A last difference concerned Fo5176 chromosome 11 of 3.09 Mb and chromosome 12 of 3.02 Mb as
well as Fo47 core chromosome 5 of 4.52 Mb.
A 5' minor portion (totalling 1.23 Mb) of Fo5176 chromosome 11 was syntenic to
a 3' minor portion (1.13 Mb) of Fo47 core chromosome 5, and
a major 5' portion of Fo47 core chromosome 5 was syntenic to Fo5176 chromosome 12.
The major portion of Fo5176 chromosome 11 was supernumerary.
Fo47 core chromosomes 5 was syntenic to a single contig or scaffold
in each of the genome assemblies of isolates 160527 and UK0001.
This observation indicated that Fo5176 chromosome 11 was composed of
a core chromosomal region and a supernumerary one,
and that Fo5176 chromosome 12 was a core one
(Fokkens et al. 2020).


Put together, the genome assembly of isolate Fo5176 consisted of
11 core chromosomes, 4 supernumerary chromosomes and 4 keeper chromosomes.
A keeper chromosome was composed of a core chromosomal region of at least 0.80 Mb
and a larger supernumerary region. Unlike supernumerary chromosomes, which
contained no essential genes, keeper chromosomes contained
a core chromosomal region with conserved genes, and their losses may be deleterious.
Keeper chromosomes play roles in the evolution of supernumerary chromosomes (see below).
The four keeper chromosomes resulted from translocations
involving core regions of at 0.8 Mb.
Two of the core chromosomes underwent a reciprocal translocation involving
core regions of at least 1.4 Mb.
Regions around each translocation breakpoint were confirmed by long reads.
Several Foc race 1 and 2 isolates belonged to the population containing Foc isolate Fo5176 (see below),
and these isolates all contained the same core chromosome translocation as isolate Fo5176.
Because this translocation was a genetic barrier to meiotic recombination
with other populations in the {\em F. oxysporum} complex,
this population is an example of potential new species that emerged during asexual reproduction
over the last hundreds of years.

\subsection*{Evidence: Two or more copies of a supernumerary chromosome in isolates}

Below we focus on two or more structurally different copies of a supernumerary chromosome in isolates.
For two or more types of supernumerary chromosomes in isolates, see the above Fusion subsection
on isolate Fo5176.

We estimated the copy numbers of Fol isolate D11 supernumerary chromosome 14 in 155 datasets of short reads,
with multiple datasets generated, one per sequencing run, from each Fol isolate.
To obtain a copy number estimate for supernumerary chromosome 14 in a dataset of short reads,
we mapped the short reads onto D11 supernumerary chromosome 14 as a reference
and separately onto the core chromosome carrying the $EF1-{\alpha}$ gene as a reference.
Then we computed a copy number estimate by dividing
the average read depth for supernumerary chromosome 14 by that for the core chromosome.
In 143 of the 155 datasets, the copy number estimates for D11 supernumerary chromosome 14
were in the range 0.92--1.38, and in the remaining 12, the values were 1.76--1.78.
These 12 datasets were generated in 12 sequencing runs from the same Fol isolate named DF0-62.
A manual examination of the read depths of D11 supernumerary chromosome 14 in
these 12 datasets revealed higher read depths over a 5' portion at 13.1 to 254.8 kb
and over a 3' portion at 1,128.1 to 2,066.4 kb of D11 supernumerary chromosome 14
than over the middle portion between them.
When only the 5' and 3' portions were used to estimate the copy numbers in these 12 datasets,
the estimates were in the range 2.09--2.12; when only the middle portion was used to compute
those, the estimates were in 1.16--1.19. 
These observations suggested that Fol isolate DF0-62 contained two structurally 
different copies of D11 supernumerary chromosome 14.
In addition, we found 509 common SNPs in the 12 datasets of short reads from isolate DF0-62;
231 of the 509 SNPs had both reference and alternate allele frequencies above 30\%
in one of the 12 datasets, implying that the two copies contained SNP differences as well.
The presence of two structurally different copies of a supernumerary chromosome
in an isolate suggested that these copies were acquired horizontally.

After mapping short reads from an isolate onto a reference chromosome or contig,
we calculated the total uncovered size of the isolate for the reference
by collecting all uncovered reference regions of sufficient lengths and totalling their sizes.
For this analysis, the minimum length of each uncovered reference region was set to 5 kb.
Of the 155 Fol isolates, 11 isolates had total uncovered sizes of 1,660 to 1,754 kb
for supernumerary chromosome 14 of 2,139 kb, and the other 144 isolates had their values less than 900 kb.
For D11 supernumerary contig 38 of 1,574 kb, these 11 isolates had total uncovered sizes of 962 to 1,046 kb,
and the rest each had their values less than 600 kb.
For D11 keeper contig 1 of 5,802 kb, 47 isolates had total uncovered sizes of
3,765 to 4,205 kb, and the remaining 108 isilates each had their values less than 2,500 kb.
These 47 isolates include all of the 11 isolates mentioned above.
For D11 keeper contig 6 of 2,135 kb, 83 isolates had total uncovered sizes of
913 to 1,184 kb, and the remaining 72 isolates each had their values less than 150 kb.
These 83 isolates include all of the 47 isolates.
Of the 155 isolates, 47 isolates were more distant to
the Fol D11 isolate than the rest;
for example, the SNP rates between these 47 isolate and the D11 core chromosome carrying
the EF1-alpha gene were estimated to be at least 84 times more than those for the rest.
This group of 47 distant isolates is exactly the same group of 47 isolates
identified above to have the largest total uncovered sizes for keeper contig 1.
Put together, these observations revealed that the 47 isolates
contained little supernumerary portions of D11 keeper contigs 1 or 6,
but of these 47 isolates, 36 isolates contained significant portions
of D11 supernumerary chromosome 14 and supernumerary contig 38.
In fact, the 36 isolates had total uncovered sizes
of 280 to 411 kb for D11 supernumerary chromosome 14,
and of 72 to 598 kb for D11 supernumerary contig 38.
In other words, standalone supernumerary chromosomes,
but not supernumerary portions of keeper chromosomes, were found in distant isolates,
suggesting that supernumerary chromosomes moved from one isolate
to another, but not keeper chromosomes. Note that
keeper chromosomes contained core chromosomal regions and their
horizontal transfer would result in the duplication of these core regions in isolates.

We also estimated the copy numbers of Focb TR4 UK0001 supernumerary chromosome 13 
in 18 Focb TR4 isolates.
We obtained a highest copy number value of 2.80 for isolate II-5,
and values in a range of 1.96 to 2.12 for 4 isolates JV11, Leb1.2C, JV14 and FOC.TR4-1,
and values in a range of 0.99 to 1.19 for the remaining 13 TR4 isolates.
These numbers suggested three copies of UK0001 supernumerary chromosome 13
in isolate II-5, and two copies of this supernumerary chromosome in the 4 TR4 isolates.
A manual examination of the read depths revealed that
the copies in each of these isolates were structurally different.

Race 1 isolate N2 was distant to the TR4 isolates in core chromosome,
and its population-specific subtelomere was different from
that of the TR4 isolates. But it was closer to the TR4 isolates
in Focb TR4 UK0001 supernumerary chromosome 13.
A section of this chromosome from 278.7 to 518.4 kb
was present in two copies in isolate N2, which contained 651 SNPs with
both reference and alternate allele frequencies above 30\%.
In the remaining portion (totalling 1004.9 kb) of the chromosome,
the number of SNPs between isolate N2 and TR4 isolate UK0001 was 230.
Moreover, the SNP rate between isolates N2 and UK0001 in this portion of the chromosome 13
was 26 times lower than their genome-wide SNP rate,
which is inconsistent with the expectation that
a supernumerary chromosome is not more variable than the core chromosomes.
An explanation to this inconsistency is that a recent horizontal transfer event involving
a version of supernumerary chromosome 13 occurred to an ancestor of isolates N2 or UK0001.
This transfer event was preceded or followed by a change in the subtelomere
of the supernumerary chromosome.
For isolate N2, an average read depth of 454.6 for a region of 661 bp at 3.55 to 4.21 kb
of a reference subtelomere for the TR4 isolates was obtained, indicating
a partial similarity between their subtelomeres.
Also, a retrotransposon of length 3.3 kb, which was located next
to the 5' subtelomere of isolate TR4 UK0001 supernumerary chromosome 13,
was present in 47 copies in isolate N2, in 12 copies in isolate TR4 II-5,
but was present in 0 or 1 copy in the other 17 TR4 isolates.

Some of the 18 TR4 isolates underwent changes in the subtelomeres of
their chromosomes. Those changes were located in region 1 at
0.5-3.7 kb and region 2 at 5.2-8.6 kb of a reference subtelomere,
the 5' subtelomere of Focb TR4 isolate UK0001 supernumerary chromosome 13.
For each of the 18 isolates,
the ratio of the average read depth of region 1 to that of region 2 was calculated.
For 9 of the 18 isolates, their ratios were
between 0.88 and 1.47, and for 8 of them, their ratios
were between 0.11 and 0.49, and for the last one (isolate Pak1.1A),
its ratio was 21.22. These large ratio differences suggested that
some of these isolates underwent changes in many of their
chromosome subtelomeres in the same region, during asexual reproduction.
Note that all of these 18 isolates belonged to the same asexual population.

\subsection*{Evidence: Origin of supernumerary chromosomes}

A supernumerary chromosome in one isolate may be a composition of portions of
several supernumerary chromosomes in another isolate, so a global measure of similarity
is less informative than a local measure of similarity.
Below we first identified similar supernumerary regions between Focb TR4 isolate UK0001 and
race 1 isolate 160527. Then we examined those regions between isolate 160527 and Foc race 1
isolate Fo5176, and within isolate Fo5176.
Next we confirmed the presence of supernumerary regions in 99 {\em F. oxysporum} isolates
assembled by Achari et al. (2020), with 16 of these isolates collected from natural ecosystem soil, 
by estimating the total length of similar regions
between the genome assembly of each isolate and the supernumerary genome of isolate Fo5176.
Finally we compared Fol isolates 4287 and D11.
Results from these comparisons indicated that
supernumerary chromosomes in a new population evolved from
ones in old populations.


Supernumerary contig 13 of 1,245 kb in isolate UK0001
contained 7 regions (totaling about 80\% of the contig)
that were syntenic to regions of the 3,211-kb supernumerary section
of keeper contig 2 in isolate 160527. Let the 7 regions in isolate 160527
be denoted by a, b, c, d, e, f and g in forward orientation in the 5'-to-3' order.
Then the order and orientation of the 7 syntenic regions in isolate UK0001
was A, F-, D, E-, C-, B- and G-,
where a letter in upper case denotes a region in isolate UK0001 that
was syntenic to the region denoted by the letter in lower case in isolate 160527,
and region F- denotes the reverse complement of region F.
The best gap-free matches in these 7 regions in the order in isolate UK0001
were 1/24,561, 2/52,381, 3/55,553, 0/33,736, 0/60,915, 1/20763 and 1/41,883,
where a gap-free alignment of $n$ paired nucleotides with $d$ nucleotide differences
was given in the form of $d/n$.
For comparison, a best gap-free match between the core chromosomes
of isolates 160527 and UK0001 was 61/10,946.

Isolate 160527 contained a keeper contig (contig 2) and a supernumerary contig (contig 12)
as well as 10 core contigs.
Four of the seven regions mentioned above in contig 2 of isolate 160527, regions c, d, e and f,
overlapped with the first three of the following four
regions of keeper contig 2 that were similar to supernumerary regions in isolate Fo5176.
Region 1 of 133 kb at 3,468 to 3,601 Mb was 91\% identical to a supernumerary region
of Fo5176 keeper chromosome 11 at 2,227 to 2,344 Mb.
Region 2 of 100 kb at 3,620 to 3,719 Mb was 93\% identical to a supernumerary region
of Fo5176 chromosome 11 at 2,127 to 2,225 Mb.
Region 3 of 290 kb at 3,986 to 4,275 Mb
was 89\% identical to a region of Fo5176 supernumerary chromosome 16
at 0.893 to 0.580 Mb (in reverse order denoting reverse orientation).
Region 4 of 145 kb at 5,237 to 5,381 Mb
was 86\% identical to a supernumerary region of Fo5176 keeper chromosome 10 at 1,997 to 2,153 Mb.
This example presented a case where a supernumerary region in one isolate
was composed of regions that were similar to supernumerary regions of
separate chromosomes in another isolate. 
In supernumerary contig 12 of 1.261 Mb of isolate 160527, 4 of its regions
(totalling 487 kb) were 87-90\% identical to supernumerary regions of Fo5176 keeper chromosome 4.


We examined homologous regions between supernumerary and keeper chromosomes in
isolate 5176 to shed light on the evolution of these chromosomes.
We found three sets of highly similar regions
between Fo5176 supernumerary chromosome 2 and keeper chromosome 13.
A first set contained a 5' region of 55 kb from keeper chromosome 13 that was 99\% identical to 
two regions of Fo5176 supernumerary chromosome 2 at 750 to 801 kb and at 922 to 976 kb.
A second set was a list of 6 close regions totalling 83 kb conserved at 98\% identity between
Fo5176 supernumerary chromosome 2 at 5.680 to 5.886 Mb and keeper chromosome 13 at 4.3 to 127.7 kb.
The last one was a set of 11 close regions totalling 60 kb conserved at 98\% identity
between chromosome 2 at 128.8 to 246.0 kb and chromosome 13 at 885 to 1,027 kb.
These matches indicated recent segmental duplications between
the two types of chromosomes in the Fo5176 lineage.
We also found 12 more sets of regions conserved at 87-97\% identity
between supernumerary and keeper chromosomes in isolate 5176.


We checked on the presence of supernumerary regions in the genome assemblies
of 99 isolates, with 16 of them collected from natural ecosystem soil.
A dataset of supernumerary sequence regions was prepared by
taking all supernumerary chromosomes and supernumerary portions of
all keeper chromosomes in the genome assembly of isolate 5176.
The total lengths of similar regions of at least 5 kb between
the supernumerary sequence regions of isolate 5176 and each of
the genome assemblies of 99 {\em F. oxysporum} isolates
ranged from 26.7 to 546.8 kb with a mean value of 186.5 kb.
This observation supported the hypothesis that all {\em F. oxysporum} isolates contained
supernumerary chromosomes or regions.

The supernumerary portion of contig 1 in Fol race 3 isolate D11
contained regions that were syntenic to supernumerary contigs 4, 18, 47, 65
in the genome assembly of Fol race 2 isolate 4287, but
the rest of it, a 1,251-kb supernumerary region at 1,124-2,375 kb,
had no long syntenic matches to any part of the 4287 genome assembly.
This supernumerary region contained long syntenic matches to a genome assembly of 
{\em F. oxysporum} isolate ISS-F4, with a best gap-free match of 12.7 kb at
a percent difference of 0.01\%.
As a comparison, a best gap-free core match of 13.5 kb between isolates D11 and ISS-F4
had a percent difference of 0.18\%, which was 18 times that of the best supernumerary match.
Note that isolates D11 and 4287 were highly similar in core chromosome;
their best gap-free core match was of length 107 kb with no differences.
This example revealed a case where a supernumerary chromosome in isolate D11 contained
two kinds of regions that originated from supernumerary chromosomes in different lineages
whose core chromosomes were more distant.
Transfer, not loss, is best at explaining this observation.
However, the comparison of the genome assemblies of isolates D11 and 4287
revealed many inter-chromosomal rearrangements between their supernumerary chromosomes;
the two isolates were in the same population (see below).

\subsection*{Evidence: Examples of populations}


The asexual subpopulation containing Focb TR4 isolate UK0001 as an individual
was defined as a group of {\em F. oxysporum} isolates
whose chromosome ends were highly similar to the subtelomere of 9.2 kb at the ends
of chromosomes in Focb TR4 isolate UK0001,
and whose lineages to the most recent common ancestor (MRCA) were all asexual.
Besides isolate UK0001, we found 17 Focb TR4 isolates (see Materials and Methods).
Short reads from all 17 Focb TR4 isolates
covered internal core AT rich regions of the UK0001 genome assembly,
suggesting that the lineages for the 17 isolates and UK0001 to the MRCA
were all asexual. For 16 of the 17 isolates, their short reads
covered the UK001 subtelomere at average read depths from 680 to 4,371, with up to 2 SNPs.
This indicated the presence of multiple highly similar copies of this subtelomere in these
16 TR4 isolates. The other TR4 isolate, Pak1.1A, had the lowest average
read depth of 440. A manual check on this coverage file found
read depths of 0 to 2 for a 2.0-kb section of the subtelomere,
and read depths of up to 90 for a 2.4-kb section of the subtelomere, and
an average depth of 675 for the rest (totalling 4.8 kb) of the subtelomere.
This showed that the subtelomeres at the ends of the Pak1.1A chromosomes
underwent significant changes.
Thus, the 16 Focb TR4 isolates belonged to the UK0001 asexual subpopulation, but isolate Pak1.1A did not.
Note that all 17 isolates was extremely close to isolate UK0001 with a SNP rate less than 0.00001.
For example, Focb TR4 isolate Pak1.1A had an average read depth of 51.6 over the whole reference genome
with a SNP rate of 6.7e-06 to the reference isolate (UK0001).
The low SNP rates indicated that the UK001 asexual subpopulation was young,
and that the changes in the subtelomere of isolate Pak1.1A happened recently,
perhaps within decades.



Another asexual subpopulation was defined based on Focb race 1 isolate 160527 with
a subtelomere of 8.5 kb identified from its high-quality genome assembly.
We found 3 additional Focb race 1 isolates: N2, VCG0124 and VCG0125.
Isolate N2 had a SNP rate of 0.00634 with isolate 160527.
Because of a lack of read coverage by short reads from isolate N2 over internal AT rich regions
of the core chromosomes for isolate 160527, either or both lineages for isolates
160527 and N2 had undergone meiotic recombination since their split.
But their subtelomeres were still homologous;
an initial 5.6-kb section of the isolate 160527 subtelomere and the rest (totalling 2.9 kb)
were covered at average depths of 918 and 135 by short reads from isolate N2.

Isolates VCG0124 and VCG0125 were close with a SNP rate of 0.00011,
but were less close to isolate 160527, with SNP rates of 0.00148 and 0.00156, respectively.
Still, the genome assemblies of isolates VCG0125 and 160527 
contained 35 gap-free matches (with at least 99\% percent identity) of lengths 3.1 to 66.8 kb totalling 403 kb
over regions with an AT content of at least 60\%,
suggesting that the lineages for these two isolates had remained asexual since their divergence. 
However, isolates VCG0124 and VCG0125 contained different types of subtelomeres.
Short reads from isolate VCG0125 revealed that
the subtelomere for isolate VCG0125 was globally similar to that of isolate 160527
with an average read depth of 1,448, where sharp drops in read depths
occurred in only two locations (at locally lowest depths of 67 and 349, respectively).
But mapping short reads from isolate VCG0124 and from all TR4 isolates
to the isolate 160527 subtelomere showed that 
the subtelomeres for these isolates including VCG0124 were only similar to
a short section (totalling 1.5 kb) of the isolate 160527 subtelomere;
for example, only this section was covered at an average read depth of 1,093
by short reads from isolate VCG0124 and at 879 by those from Focb TR4 isolate JV11.

On the other hand, two adjacent sections (at sizes of 2.8 and 1.8 kb) of the isolate UK0001 subtelomere
were covered at average read depths of 1,159 and 207 by short reads from isolate VCG0124
and at 911 and 344 by those from Focb TR4 isolate Pak1.1A. Note that this TR4 isolate was
the only one not belonging to the UK0001 population (see above).
Put together, the subtelomere for isolate VCG0124 was more similar to that
for isolate UK0001 than to that for isolate 160527.
This observation indicated that at least one of isolates VCG0124 and VCG0125 switched to
a different type of subtelomere since their divergence.
Focb race 1 isolate VCG0124 had a SNP rate of 0.00814 with isolate UK0001.
Short reads from isolate VCG0124 did not cover internal AT rich regions of the core UK0001 chromosomes,
suggesting that the lineage for isolate UK0001 underwent sexual reproduction after they split.
Thus, isolate VCG0124 did not belong to the UK0001 asexual subpopulation.
Note that short reads from isolates VCG0124 and VCG0125 covered, at read depths of at least 10,
0.61.8\% and 0.55.8\% of supernumerary contig 2 of isolate 160527,
and 45.9\% and 17.1\% of supernumerary contig 13 of isolate UK0001.
The two reference isolates, UK0001 and 160527, were more distant, with a SNP rate of 0.00959.
Although isolate VCG0124 did not belong to 
the population that included the asexual subpopulation with Focb TR4 isolate UK001,
because of a lack of global subtelomere similarity, the local subtelomere similarity
indicated a common ancestor for the portions of the subtelomeres for isolates VCG0124
and UK0001.


A third asexual subpopulation was defined based on Foc race 1 isolate Fo5176 with
a subtelomere of 9.8 kb identified from its high-quality genome assembly.
We also selected Foc race 1 isolates Cong1-1 and IVC-1, 
and Foc race 2 isolates 54008 and 58385, based on the availability of 
genome assemblies or datasets of short reads.
The SNP rate between each pair of race 1 isolates was less than 0.00003,
that between the two race 2 isolates was 0.00009,
and those between the race 1 and race 2 isolates were between 0.00006 to 0.00010.
The average read depth for the dataset of short reads from isolate IVC-1
on the Fo5176 genome assembly as a reference was 49, that from isolate 58385 was 64, and
that from isolate Cong1-1 HS1 was 42.
(The same values for the read depths for the three datasets
were obtained on the Cong1-1 genome assembly as a reference.)
The percent coverage for the AT rich region of the race 1 Fo5176 genome assembly
was 80\% for the dataset of short reads from race 1 isolate IVC-1,
94\% for that from race 1 isolate Cong1-1 HS1,
and 91\% for that from race 2 isolate 58385.
The values for those of the race 2 54008 genome assembly
from the three datasets were 73\%, 80\% and 96\%.
These numbers reflected the mode of asexual reproduction in the lineages of
these races 1 and 2 isolates up to their MRCA.
The average read depths on the 9.8-kb Fo5176 subtelomere by
the short reads from the three isolates were 3,250, 1,627 and 2,903,
indicating multiple copies of the subtelomere the genomes of these isolates.
Thus, these races 1 and 2 isolates belonged to the Fo5176 population.


A fourth population was centered on Fol isolate D11 with a high-quality genome assembly.
After mapping short reads from isolate Fol 4287 onto the genome assembly
of Fol isolate D11, we found that 57.6\% of 974 kb of AT rich regions in
the D11 genome assembly was covered by short reads from isolate Fol 4287,
suggesting that Fol isolates 4287 and D11 were in the same asexual population.
Both isolates carried nearly identical subtelomeres of 10.7 kp
(with only a few short indel differences) at the ends of their chromosomes.
Moreover, of 13 more Fol isolates, 9 were also in the D11 asexual population
based on their percent coverage of the D11 AT rich regions in a range of 32.8\% to 76.3\%,
but the other 4 were not in the population because of their low coverage in a range of 0.5\% to 4\%.
The names of these 9 Fol isolates were
14844 (M1943), 5397, CA92.95, DF0-40, DF0-41, IPO1530/B1, LSU-3, LSU-7, WCS852/E241,
and those of the other 4 were DF0-23, DF0-38, DF0-62, MN-14.
Except isolate MN-14, 
the isolates also had highly similarly subtelomere sequences in high copy numbers.
Thus, except isolate MN-14, the isolates belonged to the D11 population.
A detailed description of the analysis of th 155 runs from these 13 isolates is given below.

The Fol D11 population was a well-studied population, with more than 100
datasets of short reads at NCBI.
To check if isolates in the population carried multiple copies of the subtelomere
at chromosome ends of the Fol D11 reference genome assembly,
we mapped datasets of short reads from 155 runs of 13 Fol isolates (11-12 runs per isolate)
onto supernumerary chromosome 14
and contig 38 (as well as keeper contigs 1 and 6 each with both supernumerary and core regions)
of Fol isolate D11 as a reference.
Of the 155 runs, 144 runs each carried a significant portion (more than 30\%) of
each of supernumerary chromosome 14 and contig 38, and 
carried multiple copies of the subtelomere based on read coverage.
None of the remaining 11 runs carried a significant portion of
either of supernumerary chromosome 14 and contig 38, or carried any copies of the subtelomere.
Moreover, none of these 11 runs carried a significant portion of
the supernumerary region of either keeper contig 1 or contig 6.
Of the 144 runs carrying multiple copies of the subtelomere,
108 runs carried a significant portion of the supernumerary region of keeper contig 1,
and 72 runs carried that of keeper contig 6.
By defining the Fol population as a group of isolates with the Fol subtelomere
at the ends of their chromosomes, we classified the 144 runs as
belonging to the Fol population. Thus, all isolates in the population carried
a supernumerary region of each of supernumerary chromosome 14 and contig 38,
but some did not carry any of the supernumerary regions of keeper contigs 1 and 6.



\section*{Materials and Methods}

We obtained the genome assemblies of the following isolates
(by their GenBank assembly accessions) from GenBank
at National Center for Biotechnology Information (NCBI):
{\em F. oxysporum} f.sp. {\em conglutinans} (Foc) race 1 isolate Fo5176 (GCA\_014154955.1),
{\em F. oxysporum} f.sp. {\em conglutinans} (Foc) race 1 isolate IVC-1 (GCA\_014839635.1),
{\em F. oxysporum} f.sp. {\em conglutinans} (Foc) race 1 isolate Cong1-1 (GCA\_018894095.1),
{\em F. oxysporum} f.sp. {\em conglutinans} (Foc) race 2 isolate 54008 (GCA\_000260215.2),
{\em F. oxysporum} f.sp. {\em conglutinans} (Foc) race 2 isolate 58385 (GCA\_002711385.1),
{\em F. oxysporum} f.sp. {\em conglutinans} (Foc) isolate FGL03-6 (GCA\_002711405.2),
{\em F. oxysporum} f.sp. {\em cubense} (Focb) race 1 isolate 160527 (GCA\_005930515.1),
{\em F. oxysporum} f.sp. {\em cubense} (Focb) tropical race 4 (TR4) isolate UK0001 (GCA\_007994515.1),
{\em F. oxysporum} f.sp. {\em lycopersici} (Fol) race 3 isolate D11 (GCA\_003977725.1),
{\em F. oxysporum} f.sp. {\em lycopersici} (Fol) race 2 isolate 4287 (GCA\_001703175.2),
{\em F. oxysporum} f.sp. {\em melongenae} (Fom) isolate 14004 (GCA\_001888865.1),
{\em F. oxysporum} f.sp. {\em melonis} (Fom) isolate 26406 Fom001 (GCA\_002318975.1),
{\em F. oxysporum} f.sp. {\em radicis-cucumerinum} (Forc) isolate Forc016 (GCA\_001702695.2),
{\em F. oxysporum} isolate FISS-F4 (GCA\_004292535.1),
{\em F. oxysporum} isolate Fo47 (GCA\_013085055.1), and
99 {\em F. oxysporum} isolates whose genome assemblies produced by Achari et al. (2020).

We also obtained the datasets of short reads for
the following isolates (by their SRA accessions)
from Sequence Read Archive (SRA) at NCBI:

2 {\em F. oxysporum} f.sp. {\em conglutinans} (Foc) race 1 isolates
IVC-1 (SRR11823424), Cong1-1 HS1 (SRR12709665);

1 {\em F. oxysporum} f.sp. {\em conglutinans} (Foc) race 2 isolate
58385 (SRR8640621);

3 {\em F. oxysporum} f.sp. {\em cubense} (Focb) race 1 isolates
N2 (SRR550150), VCG0124 (SRR13311630), VCG0125 (SRR13311629);

18 {\em F. oxysporum} f.sp. {\em cubense} (Focb) TR4 isolates
Col4 (SRR10125423), Col17 (SRR10747097), Col2 (SRR10103605),
FOC.TR4-1 (SRR10054450), FOC.TR4-5 (SRR10054449),
Hainan.B2 (SRR550152), II-5 (SRR10054446), JV11 (SRR7226881),
JV14 (SRR10054448), La-2 (SRR7226878), Leb1.2C (SRR7226880),
My-1 (SRR7226877), Pak1.1A (SRR7226883), Phi2.6C (SRR7226882),
S1B8 (SRR10054447), VCG01213/16 (SRR13311628), Vn-2 (SRR7226879),
UK0001 (SRR9733598);


13 {\em F. oxysporum} f.sp. {\em lycopersici} (Fol) isolates with a total of 155 runs (11-12 runs per isolate)
CA92/95 (12 runs: SRR307095, SRR307102, SRR307126, SRR307129, SRR307254, SRR307274,
SRR307299, SRR307331, SRR307235, SRR307240, SRR307246, SRR307276),
LSU-3 (12 runs: SRR307087, SRR307089, SRR307093, SRR307118, SRR307233, SRR307234,
SRR307252, SRR307267, SRR307256, SRR307271, SRR307307, SRR307328),
LSU-7 (12 runs: SRR307111, SRR307237, SRR307239, SRR307249, SRR307268, SRR307327,
SRR307341, SRR307345, SRR307261, SRR307284, SRR307323, SRR307347),
IPO1530/B1 (12 runs: SRR307080, SRR307094, SRR307098, SRR307103, SRR307104, SRR307288,
SRR307292, SRR307298, SRR307291, SRR307296, SRR307301, SRR307312),
DF0-41 (12 runs: SRR307244, SRR307265, SRR307302, SRR307334, SRR307108, SRR307112,
SRR307121, SRR307124, SRR307242, SRR307253, SRR307311, SRR307325),
WCS852/E241 (12 runs: SRR307084, SRR307241, SRR307264, SRR307273, SRR307282, SRR307286,
SRR307303, SRR307348, SRR307272, SRR307280); SRR307315 SRR307324),
14844(M1943) (12 runs: SRR307081, SRR307116, SRR307117, SRR307119, SRR307269, SRR307332,
SRR307342, SRR307346, SRR307236, SRR307255, SRR307293, SRR307316),
5397 (12 runs: SRR307083, SRR307085, SRR307109, SRR307110, SRR307120, SRR307260,
SRR307314, SRR307319, SRR307247, SRR307277, SRR307310, SRR307333),
DF0-40 (12 runs: SRR307088, SRR307099, SRR307105, SRR307125, SRR307127, SRR307245,
SRR307309, SRR307339, SRR307279, SRR307313, SRR307321, SRR307344),
DF0-23 (12 runs: SRR307106, SRR307107, SRR307113, SRR307115, SRR307123, SRR307238,
SRR307257, SRR307266, SRR307295, SRR307297, SRR307322, SRR307336),
DF0-38 (12 runs: SRR307086, SRR307090, SRR307091, SRR307092, SRR307122, SRR307250,
SRR307262, SRR307278, SRR307281, SRR307306, SRR307320, SRR307326),
DF0-62 (12 runs: SRR307100, SRR307101, SRR307128, SRR307130, SRR307248, SRR307258,
SRR307259, SRR307270, SRR307275, SRR307300, SRR307317, SRR307343),
MN-14 (11 runs: SRR307082, SRR307096, SRR307097, SRR307243, SRR307263, SRR307285,
SRR307290, SRR307318, SRR307335, SRR307338, SRR307340).

Each paired-end dataset of short reads was represented by a pair of Fastq files
in compressed format. For example,
the names of the two files for a dataset with SRA accession SRR3139043
were SRR3139043\_1.fastq.gz and SRR3139043\_2.fastq.gz.
To process many datasets in a batch mode, their pairs of files
were placed in a data directory. A Linux shell script was written
to go through each pair of files and to call a Perl script
to map the files of short reads onto a reference genome assembly. 
The path of the data directory was included in the Perl script.
The Perl script takes as input an SRA accession number and the name of 
a fasta file containing the reference genome assembly in the working directory.
Then it calls Bowtie2 (Langmead and Salzberg 2012)
to map the two files of short reads onto
the reference genome assembly, generating an alignment output file in BAM format.
Next the Perl script calls Picard to transform the BAM alignment file
and calls GATK with command option HaplotypeCaller to produce a file (in VCF format)
of SNPs and indels between the short reads and the reference.
After that, it calls Bedtools (Quinlan and Hall 2010) with command option genomecov to
report a file (whose name ends in `.cov') of reads depths at each reference genome position
and to report a file of reference regions with zero coverage in BedGraph format.
Finally, the Perl script calls a custom AWK (named `z.cov.awk') script to calculate
the average read depth for the reference from the file of reads depths
and calls another custom AWK script to calculate a total size of reference regions
of at least 5 kb with zero coverage for each chromosome or contig in the reference genome assembly
from the BedGraph file.

The z.cov.awk AWK script for calculating the average read depth for the reference
from the .cov file was based on the following algorithm.
The reference can be a whole genome assembly or a chromosome.
If the reference is a chromosome with a subtelomere element at an end,
then the subtelomere element would be covered at high depths by short reads from
the multiple copies of the subtelomere element in the genome.
These high depths of coverage would inflate the average depth of the chromosome.
To address this problem, we used the standard formula for calculating
the original mean and standard deviation of the read depths for the reference.
Then we calculated a revised mean by using only those reference positions
whose read depths were not more than the original mean plus three times
the standard deviation. The revised mean was reported in a file whose name ends in `.average2'
by the z.cov.awk AWK script from the the .cov file.

To determine if an isolate with a dataset of short reads belong to the asexual
population of a reference isolate with a genome assembly in a fasta sequence file,
the fasta sequence file was processed by a custom AWK script to produce an output file
(whose name ends in `.ATareas') of AT rich regions of sizes at least 2 kb,
where each region is composed of multiple consecutive lines of the fasta sequence file such
that the AT content of each line is at least 65\%. 
Then another custom AWK script was written to calculate the percentage of
the AT rich regions that was covered by the dataset of short reads at
read depths of at least 5, where the AWK script takes as input
the .ATareas file and the .cov file.
The isolate belongs to the asexual population of the reference isolate
if the percentage of the reference AT rich regions covered by
short reads from the isolate was above a cutoff, say, 10\%.
In addition, the SNP rate between the short read isolate and the reference isolate
was calculated as the number of SNPs with a read depth of at least 10
(given in the DP field in the VCF file)
and a quality value of at least 80 (given in column 6 of the VCF file)
divided by the number of reference positions with a read depth of at least 10
(obtained by using the .cov file).


Candidate genes in a genome assembly were found by AUGUSTUS
(Stanke and Waack 2003).
Functional annotation of predicted protein sequences
wer performed by HMMER (Finn et al. 2011).
Gap-free matches within a genome assembly were computed by
the DDS2 program (Huang et al. 2004).
The output from the program was filtered by an AWK script
to select gap-free matches of at least 5 kb in length and at least 99\% in identity.
This procedure was applied to the genome assemblies of isolates Fo47 and Foc Fo5176. 

A genome assembly may still contain
subtelomeres that were inactivated by RIP during the last sexual cycle.
To find such subtelomeres in the genome assembly,
we used the AAT package (Huang et al. 1997) to search the genome assembly
for matches to an ATP-dependent DNA helicase hus2/rqh1 of 1,428 residues
(accession: KAG7001869.1) encoded in isolate Fo5176 supernumerary chromosome 18.
Genome sequences that were similar to the DNA helicase were potential subtelomeres
as they encoded a helicase, a signature of subtelomeres.
The DNA-protein alignment produced by AAT was used to count
the number of stop codons in the reading frame of the DNA sequence,
where each stop codon in the reading frame was marked with three stars.
Note that some of the matches might be intact subtelomeres that
contained no stop codons in their reading frames.
For example, the sequences of the 3' 30-kb ends of isolate Forc Forc016 chromosomes 4 and 10
were similar and contained intact subtelomeres adjacent to inactivated subtelomeres,
as indicated by an alignment of these sequences produced by
the SIM program (Huang and Miller 1991), where all 434 base mismatches in
the alignment were located in the inactivated subtelomeres.
This also happened to those of Forc016 chromosomes 4 and contig 7.

All isolates with sufficiently high percent coverage of the AT rich regions
in a reference isolate had low SNP rates with the reference isolate.
Conversely, all isolates with high SNP rates with a reference isolate
had low high percent coverage of the AT rich regions in the reference isolate.
This association means that a measure of SNP rate is useful
in predicting the mode of reproduction in isolates.
The distribution of SNP rates between isolates can be used
to predict how long asexual reproduction lasts.
Assume that 10 SNPs occurred per genome per year.
Then two isolates with thousands of SNPs were estimated
to diverge hundreds of years ago.
If a sample of isolates from a population were estimated
to have hundreds of SNPs among them, then the population
was estimated to have existed for decades.

\section*{Discussion}

Complementary mechanisms of gene flow play important roles
in speciation and species maintenance.
Nonrecombining genome regions, horizontal DNA transfer and transposon proliferation
are critical factors in speciation,
while recombining genome regions, vertical DNA transfer and transposon control
are critical factors in maintaining species.
In populations of organisms that reproduce asexually and sexually,
asexual reproduction, which is different from clonal reproduction,
is an efficient mode by which speciation occurs,
and sexual reproduction is an effective mode by which species are maintained.
In the {\em F. oxysporum} complex, a large number of populations arose
for diverse plant and animal hosts during asexual reproduction,
where extensive structural variation among isolates in the same asexual population 
overwhelmed nucleotide substitution variation among those isolates, and
some populations contained core chromosome rearrangements.
In homothallic fungi, the recombination between 
chromosomes from different nuclei in the same individual plays
an important role in controlling transposons and maintaining chromosome core number and structure.
In plants, polyploid speciation occurs during asexual reproduction in one or two generations
and it may take many thousands of generations to new species to occur during sexual reproduction
(Rieseberg and Willis 2007).
Asexual reproduction is likely to play an important role in polyploid speciation
(Herben et al. 2017).
The rareness of ancient asexuals supports the claim
that asexual reproduction is not an effective mode by which
species are maintained for a long time.
Bacteria is an exception because they contain single circular
chromosomes, without separate chromosomes. 
This suggests that recombination is the most effective
mechanism for maintaining chromosome number and structure,
which plays a key role in maintaining species.

The {\em F. oxysporum} complex illustrates difficulty in defining species.
On the other hand, the presence of gene flow between asexual populations
in this complex offers support for the definition of species
as groups of populations with gene flow within and between populations
at population-specific rates, where complementary mechanisms of gene flow
could operate in populations.
Lack of extensive variation in core chromosome number and structure among
isolates in the complex over 10,000 years suggests that it was important
to maintain core chromosome number and structure among isolates
in the complex through sexual reproduction.
Thus, the maintenance of core chromosome number and structure as well as
the presence of gene flow may be necessary properties of any eukaryotic species,
whether they are sexual or not.

The emergence of new species and the maintenance of existing species
may be in conflict with each other, because new species may compete
with existing species for resources. We discuss how this competition
is addressed in general and in particular in the {\em F. oxysporum} complex.
According to the complementary theory, for populations of organisms
that reproduce sexually and asexually, new species emerge during asexual reproduction.
It is known that organisms reproduce asexually when resources are abundant and
switch to sex when resources are limited.
Put together, new species emerge when resources are abundant, making the competition
with existing species less likely to occur.
In the {\em F. oxysporum} complex, populations contained molecular
signatures in the form of population-specific subtelomeric palindromes at
the end of each chromosome. Thus, populations can be distinguished
from each other as far as gene flow is concerned. Also, populations had specific plant hosts.
Thus, new populations are less likely to compete with existing populations for resources.
Moreover, the availability of new resources causes new populations to emerge.

Plants and their fungal pathogens use different forms of gene duplication,
an important evolutionary strategy, to compete against each other.
Plants can multiply their large genomes, while fungi cannot carry large genomes.
The {\em F. oxysporum} complex used a novel form of gene duplication.
The complex contained a large number of populations for different hosts,
with each population carrying specific supernumerary chromosomes, while
the core chromosomes in all populations were similar and free of duplicated genes.
Novel supernumerary chromosomes emerged in existing populations
and were duplicated in new populations by horizontal transfer during asexual reproduction.
This discussion reveals that horizontal transfer plays a more important role than
vertical transfer in the emergence of new populations in this complex.
It remains unclear whether the horizontal transfer of B chromosomes is important
for new plant species to emerge.

In population genetics, evolution is defined as a change in the frequency
of gene variants. However, our analysis of genome data in the {\em F. oxysporum} complex
showed that evolution also involves a change in the frequency
of chromosome structural variants as well as horizontal chromosome transfer.
These new types of variation and transfer are important in speciation.
Also, asexual reproduction is not only different clonal reproduction,
but also is an efficient mode by which speciation occurs.
Deleterious mutations during asexual reproduction could
take the form of transposon proliferation.
The new theory calls for new mathematical models of and experimental approaches to population genetics
to address evolutionary forces acting on non-Mendelian genetic variation,
which enables new species to emerge.

\section*{Additional Information and Declarations}

\subsection*{Competing Interests}

The author is interested in exploring the potential of the genomic insights in industrial applications.

\subsection*{Author Contributions}

Xiaoqiu Huang conceived and designed the experiments, performed the experiments,
analyzed the data, contributed reagents/materials/analysis tools, wrote the paper,
reviewed drafts of the paper.

\subsection*{Data and Code Availability}

All sequencing and genomic data were downloaded from NCBI (see Materials and Methods for their
accession numbers).
Scripts used to process the data and sample output from them
are available on the Open Science Framework at https://osf.io/86y5r/

\subsection*{Funding}

This work was supported by Iowa State University.
The funders had no role in study design, data collection and analysis,
decision to publish, or preparation of the manuscript.

\section*{Acknowledgments}

The analysis of genome data was performed on high performance computing clusters
managed by the Research IT team in College of Liberal Arts and Sciences
at Iowa State University.

\end{document}